\documentclass[twocolumn]{elsart}
\usepackage[dvips]{graphicx}

\usepackage{amsmath,amssymb,amsfonts}
\usepackage{dcolumn}
\usepackage{bm}
\usepackage[usenames]{color}
\usepackage[normalem]{ulem}

\begin{document}
\begin{frontmatter}

\title{Interface induced states at the boundary between a 3D topological
insulator Bi$_2$Se$_3$ and a ferromagnetic insulator EuS}

\author[ispms,tsu,dipc]{S.V. Eremeev},
\author[Kurch,tsu,dipc]{V.N. Men'shov},
\author[Kurch,Prokh,tsu]{V.V. Tugushev},
\author[dipc,ehu,tsu]{E.V. Chulkov\corauthref{cor1}}

\address[ispms]{Institute of Strength Physics and Materials Science, 634021 Tomsk, Russia}
\address[tsu]{Tomsk State University, 634050 Tomsk, Russia}
\address[Kurch]{NRC Kurchatov Institute, 123182 Moscow, Russia}
\address[Prokh]{A.M. Prokhorov General Physics Institute, 119991 Moscow, Russia}
\address[dipc]{Donostia Internation Physics Center (DIPC), 20018 San Sebasti\'an/Donostia, Spain}
\address[ehu]{Departmento de F\'isica de Materiales UPV/EHU, CFM-MPC, and Centro
Mixto CSIC-UPV/EHU, 20080 San Sebasti\'an/Donostia, Spain}

\corauth[cor1]{E-mail: waptctce@ehu.es}

\begin{abstract}
By means of relativistic density functional theory (DFT)
calculations we study electron band structure of the topological
insulator (TI) Bi$_2$Se$_3$  thin films deposited on the
ferromagnetic insulator (FMI) EuS substrate. In the Bi$_2$Se$_3$/EuS
heterostructure, the gap opened in the spectrum of the topological
state has a hybridization character and is shown to be controlled by
the Bi$_2$Se$_3$ film thickness, while magnetic contribution to the
gap is negligibly small. We also analyzed the effect of  Eu doping
on the magnetization of the Bi$_2$Se$_3$ film and demonstrated that
the Eu impurity induces magnetic moments on neighboring Se and Bi
atoms an order of magnitude larger than the substrate-induced
moments. Recent magnetic and magneto-transport measurements in
EuS/Bi$_2$Se$_3$ heterostructure are  discussed.

\end{abstract}

\begin{keyword}
Topological insulators; Magnetic proximity effect

\PACS{73.20.-r, 75.70.Tj, 85.75.-d}

\end{keyword}
\end{frontmatter}

\section{Introduction}

Since the discovery of three dimensional TIs \cite{Hasan_RMP2010},
magnetic proximity effect (MPE) at the TI/FMI interface has been
considered as  an effective method to  break time-reversal  symmetry
of topological states  by introducing an exchange field from FMI
into TI. Such physical phenomena as quantum anomalous Hall effect
and topological magneto-electrical effect critically rely on finding
a way  to open up a magnetic gap at the Dirac point of the
topological states. In our recent works
\cite{Eremeev_proximity,Menshov_proximity}, where the thorough
studies of MPE  in  TI containing heterostructures have been
presented, it was argued that  the physics of MPE is intricate
enough and cannot be naively  described in the terms of
two-dimensional Dirac-like Hamiltonian with an added massive term.
On the other hand, intriguing peculiarities in magnetic and
magneto-transport properties of the EuS/Bi$_2$Se$_3$
heterostructures have been recently reported in Refs.
\cite{Wei,Yang}. It was shown the emergence of a ferromagnetic phase
in TI with a significant saturation magnetization \cite{Wei}; below
the Curie temperature, unusual negative magnetoresistance was
observed for the Bi$_2$Se$_3$ films thinner than $\sim4$ quintuple
layers (QL) deposited on EuS \cite{Yang}. Motivated by these
important experimental results, below we study the electron band
structure of the EuS/Bi$_2$Se$_3$ interface within the DFT method.
We focus on the spin polarization of interface states and the spin
density redistribution between FMI (EuS) and TI (Bi$_2$Se$_3$)
components of the heterostructure.

\section{Computational details}

For structural optimization and electronic band calculations we use
the Vienna Ab Initio Simulation Package (VASP) \cite{VASP2} with
generalized gradient approximation (GGA) to the exchange correlation
potential. The interaction between the ion cores and valence
electrons was described by the projector augmented-wave method
\cite{PAW2}. Relativistic effects, including spin-orbit coupling,
were fully taken into account. To describe the highly correlated
Eu-4$f$ electrons we include the correlation effects within the
GGA+$U$ method \cite{U}. The values of $U$ and $J$ parameters were
taken to be of 7.4 and 1.1 eV, respectively \cite{Larson}.

\begin{figure}
  \includegraphics[width=0.95\columnwidth]{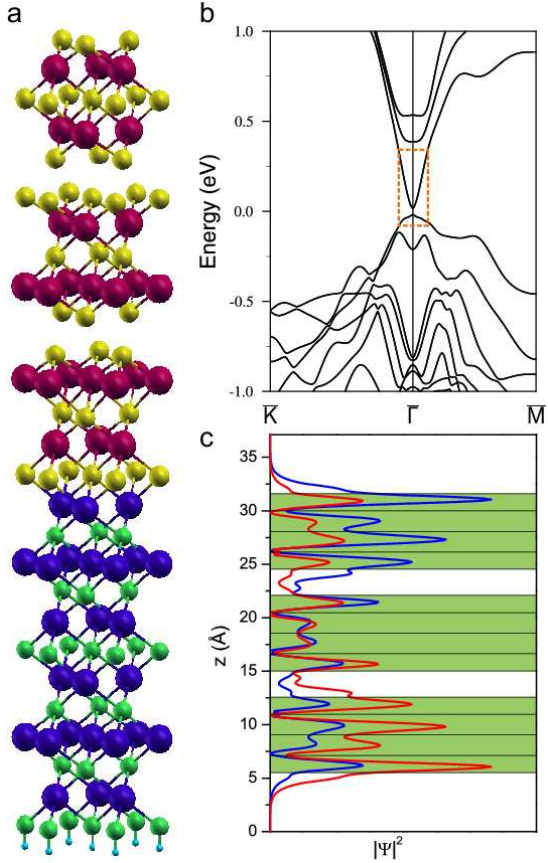}
    \caption{(Color online) (a) Bi$_2$Se$_3$(3QL)/EuS
    heterostructure. Darkred and yellow balls indicate Bi and Se atoms, respectively; Darkblue (Eu) and green (S) balls are atoms of the EuS substrate,
    lightblue dots indicate passivation hydrogen layer on the free S-terminated surface of EuS. (b) Electronic spectrum of the
    free-standing 3QL-thick film of Bi$_2$Se$_3$. Gapped topological state is framed by orange dashed-line rectangle. (c) Charge density distribution along the axis $z$ for double-degenerate gapped state
    at the $\bar\Gamma$ point; horizontal lines show position of atomic layers, shaded areas mark QLs.}
\label{fig1}
\end{figure}

To simulate the interface between FMI EuS and TI Bi$_2$Se$_3$ thin
film, the lattice constant of the rocksalt EuS in the hexagonal
(111) plane is fixed to that of Bi$_2$Se$_3$ which is $\approx 1.8$
\% smaller than the experimental lattice parameter of EuS. The
EuS(111) substrate was modeled by a slab consisting of 6 EuS
bilayers. Eu-terminated side of this slab was chosen as interfacial
plane while dangling bonds at the S-terminated side of the slab were
passivated by hydrogen atoms. The atomic positions within the four
near-interface atomic layers of EuS slab and of Bi$_2$Se$_3$ thin
film were optimized. A vacuum spacer of $\sim 15$\AA\ was included
to ensure negligible interaction between opposite sides of the
structure. The optimized structure containing interface between
EuS(111) slab and 3 QL-thick film of Bi$_2$Se$_3$ is shown in
Fig.~\ref{fig1}(a). Since selenium and sulfur are isoelectronic
elements, the interfacial (first) QL forms a strong bonding with EuS
substrate. The equilibrium Se-Eu bondlength at the interface is
3.043 \AA\ that is only slightly larger than S-Eu bond in EuS bulk
(2.980 \AA).

\section{Results}

Three dimensional TI (e.g. Bi$_2$Se$_3$) promote the formation the
topologically protected metallic states at the surface
\cite{Hasan_RMP2010} while when TI is reduced to a film of a few QL
thickness, the topological surface states became gapped
\cite{Zhang_NatPhys2010,Landolt,Silkin}. The electronic spectrum of
free-standing 3QL Bi$_2$Se$_3$ film shown in Fig.~\ref{fig1}(b)
demonstrates the gap of 36.4 meV at $\bar\Gamma$. The gap results
from spatial overlap of the states from the opposite surfaces of the
thin film (Fig.~\ref{fig1}(c)), and its size is determined by the
film thickness.

\begin{figure}
  \includegraphics[width=\columnwidth]{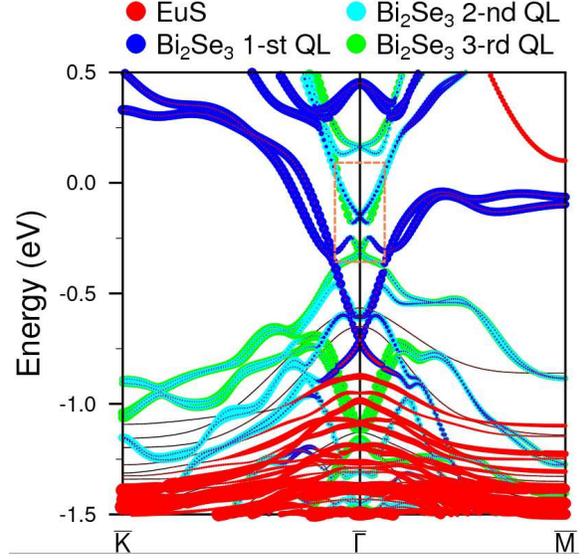}
    \caption{(Color online) Electronic structure of Bi$_2$Se$_3$(3QL)/EuS.
  Size of red circles corresponds to the weight of the states in the 5 atomic layers of EuS
  adjacent to the interface; darkblue, lightblue and green circles denote weight of the states in
  the first (interfacial), second, and third (outermost) QLs of Bi$_2$Se$_3$, respectively.}
 \label{fig2}
\end{figure}

\begin{figure}
  \includegraphics[width=\columnwidth]{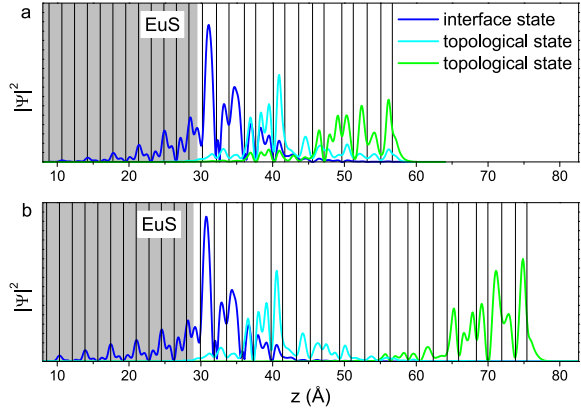}
    \caption{(Color online) Spatial localization of the ordinary interfacial (darkblue line)
    and topological states localized at inner (lightblue line) and outer (green line) side of the Bi$_2$Se$_3$ film of 3QL (a) and 5QL (b) thickness.}
 \label{fig3}
\end{figure}

The electronic structure of the 3QL Bi$_2$Se$_3$ film deposited on
EuS slab is shown in Fig.~\ref{fig2}. The spectrum demonstrates
similarities with that for a Bi$_2$Se$_3$/MnSe heterostructure
\cite{Eremeev_proximity}. As in the latter case at the boundary
between Bi$_2$Se$_3$ and EuS the topological and ordinary
interfacial states arise \cite{Eremeev_proximity,Menshov_proximity}.
It happens owing to modification of the electrostatic potential in
the near-interface region caused by accumulation of $\sim 0.6$
electron within the interfacial QL. The modified potential produces
trapping of the state localized in the first QL and its shift to the
lower energies. In contrast to the Bi$_2$Se$_3$/MnSe case, where the
ordinary interfacial state has a gap of 56 meV, the occupied
interfacial state in Bi$_2$Se$_3$/EuS, lying at energy of $-0.71$ eV
at $\bar\Gamma$, has a tiny $\bar\Gamma$-gap of 7 meV. The small
magnetic splitting is explained by the fact that the strongly
localized Eu-$f$ orbitals do not contribute to the interfacial
state. This is responsible for negligible induced magnetization of
the Bi$_2$Se$_3$ film provided via MPE. So, the magnetic moments on
the Se atoms, closest to the interface plane, and on the next Bi
atoms are -0.023 and +0.012 $\mu_{\mathrm B}$, respectively, and
they are an order of magnitude smaller on the deeper atoms of the
first QL. Like in the Bi$_2$Se$_3$/MnSe case the occupied
interfacial state has the maximal localization near the interface
plane (Fig.~\ref{fig3}a, darkblue line), decaying into the EuS and
Bi$_2$Se$_3$. As a function of momentum it extends through the bulk
gap, crossing the Fermi level.

The gapped topological state become nondegenerate acquiring a
Rashba-like spin splitting since the inversion symmetry in
Bi$_2$Se$_3$ film is broken due to the interface
\cite{Zhang_NatPhys2010,Landolt}. The spatial localization of the
topological state mainly resided at the outermost QL
(Fig.~\ref{fig3}a, green line) is almost unchanged as compared to
that in the free-standing Bi$_2$Se$_3$ film (Fig.~\ref{fig1}(c)). In
contrast, spatial localization of the state resided on the opposite
side of the Bi$_2$Se$_3$ film, adjacent to EuS substrate, is
strongly modified (Fig.~\ref{fig3}a, lightblue line). This state is
relocated away from the interface plane and its probability maximum
lies in the second QL. The similar behavior was found for the
topological Dirac state in Bi$_2$Se$_3$/MnSe
\cite{Eremeev_proximity}. The energy gap in topological state of the
3QL-thick film deposited on EuS substrate increases with respect to
the free-standing Bi$_2$Se$_3$ film, up to 64.6 meV, due to the
relocation of the topological state away from the interface plane
rather than by the MPE which is too small. This explanation is
consistent with the fact that in the 2QL-thick Bi$_2$Se$_3$ film
with broken inversion symmetry the gap between Rashba-like
spin-split topological states was found to be 97.3 meV
\cite{Landolt}.

\begin{figure}
  \includegraphics[width=\columnwidth]{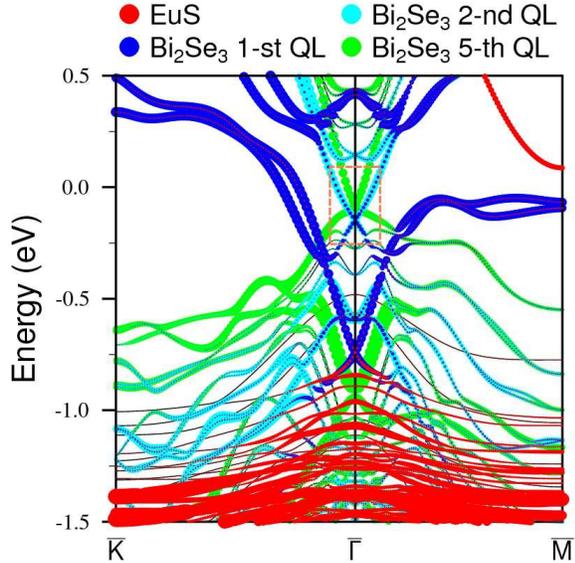}
    \caption{(Color online) Electronic structure of Bi$_2$Se$_3$(5QL)/EuS.}
 \label{fig4}
\end{figure}

In order to confirm that the gap in the topological state in
Bi$_2$Se$_3$/EuS system is controlled by the thickness of the
Bi$_2$Se$_3$ film only we increased its thickness up to 5QLs. As one
can see in Fig.~\ref{fig3}b, the spatial distribution of the
ordinary interfacial state (darkblue line) is not sensitive to the
increase in the film thickness. The topological state localized at
the inner side of the Bi$_2$Se$_3$ film, as in the previous case, is
relocated away from the interface plane (lightblue line). At the
same time the inner-side- and outer-side-resided (green line)
topological states become decoupled owing to decaying overlap with
the increasing film thickness. This results in formation of the
gapless topological Dirac cones at energies of $-0.15$ and $-0.08$
eV, localized at the inner and outer side of the Bi$_2$Se$_3$ film,
respectively (Fig.~\ref{fig4}). Moreover, the group velocities in
these states are almost the same and well comparable with that in a
thick Bi$_2$Se$_3$ slab.

The obtained band structure of the 5QL Bi$_2$Se$_3$ film deposited
on EuS differs from that of the Bi$_2$Se$_3$/MnSe heterostructure
\cite{Eremeev_proximity} where the Dirac state is gapped due to the
interaction with spin-polarized ordinary interface state. As is
stated above, at the ideally sharp Bi$_2$Se$_3$/EuS interface, the
MPE is insignificant. But what happens if the magnetic atoms are
introduced into the TI film? To study the possible effects of the Eu
atom doping on the magnetization of Bi$_2$Se$_3$ we use a
free-standing 1QL block with $3\times 3$ supercell where one of the
nine atoms in the outer selenium layer is substituted by an Eu atom.
The largest magnetic moments, 0.24 $\mu_{\mathrm B}$, are obtained
at the second-layer Bi atoms, closest to the introduced magnetic
impurity. The third-layer Se atoms, lying next to the magnetized Bi
atoms as well as fourth-layer Bi atom, lying beneath the Eu atom,
have smaller induced magnetic moments, 0.05 and 0.06 $\mu_{\mathrm
B}$, respectively. The calculated net magnetization induced on the
Bi and Se atoms of the supercell is found to be 1.20 $\mu_{\mathrm
B}$.

\section{Discussion and conclusion}

Results of our theoretical study shed light on the interpretation of
magnetic and magneto-transport experiments in Bi$_2$Se$_3$/EuS
heterostructures \cite{Wei,Yang}. Our calculations do not
corroborate the idea of the authors of Refs.~\cite{Wei,Yang} about a
crucial role of MPE in Bi$_2$Se$_3$/EuS heterostructures, leading to
strong spin polarization of interface electron states in
Bi$_2$Se$_3$ film and enhancement of indirect exchange coupling
between Eu local moments in EuS layer via these states.  On the
contrary, the calculations reveal an utterly small magnitude of the
induced spin polarization of carriers under ideally sharp boundary
between EuS and Bi$_2$Se$_3$. It is quite reasonable to attribute
the observed the ``interface ferromagnetism" \cite{Wei,Yang} to the
Eu atoms diffusion from the EuS into the first QL of Bi$_2$Se$_3$
during the heterostructure growth. According to our calculations,
the spin magnetization produced by the Eu impurity atom is 1.20
$\mu_{\mathrm B}$ per $3\times3$ unit cell of Bi$_2$Se$_3$.
Multiplying this quantity by the Land\'e $g$-factor of $\sim 50$
inherent in Bi$_2$Se$_3$ and presuming that 0-10 at.\% of Eu atoms
penetrate from the EuS substrate into the first QL of Bi$_2$Se$_3$,
we can estimate  the effective magnetization due to ``interface
ferromagnetism" as 7-13 $\mu_{\mathrm B}$ per Eu atom. That
corresponds very well to the saturation magnetic moment reported in
Ref.~\cite{Wei}.

We demonstrated that a weakly gapped ordinary bound state caused by
the interface potential arises in the immediate region adjacent to
the Bi$_2$Se$_3$/EuS interface. The topological state also exists in
the vicinity of the interface, however it is relocated from the
interface plane to distant atomic layers of the TI film and the
spectrum of this state is remarkably changed with the TI film
thickness. Thereby there are two distinct conducting channels near
the interface, ordinary and topological, which suggest original
character of low-temperature transport properties of the
Bi$_2$Se$_3$/EuS system, in particular, unusual transition from the
positive to negative magneto-resistance as the TI film becomes
thinner than the critical thickness ($\sim 4$ nm \cite{Yang}).
Following our calculations, it can be shown that Eu atoms diffusion
can play significant role in magneto-transport properties of the
Bi$_2$Se$_3$/EuS heterostructure, providing the spin polarization of
the ordinary bound state at the interface. At  large thicknesses of
the Bi$_2$Se$_3$ layer,  the weakly spin polarized high-mobility
topological Dirac-like state relocated from the interface  dominates
the electron  transport properties of the system and  thus defines
positive sign of magneto-resistance.  At small thicknesses of the
Bi$_2$Se$_3$ layer, when the topological state becomes  gapped due
to hybridization effect, only the strongly spin polarized (due to
above discussed ``interface ferromagnetism" effect) ordinary bound
state participates in the electron  transport and provides the
negative sign of magneto-resistance. The critical thickness of the
TI layer observed in \cite{Yang} testifies to the crossover between
contributions of topological and ordinary bound states into the
total magneto-resistance.

In summary, in this work we performed the DFT calculations to study
interfacial states in the Bi$_2$Se$_3$/EuS heterostructures. We
demonstrated that   ordinary and topological bound states co-exist
near the interface, thus defining a complex character of magnetic
and magneto-transport properties of the system.

\begin{ack}
We acknowledge partial support from the Basque Country Government,
Departamento de Educaci\'{o}n, Universidades e Investigaci\'{o}n
(Grant No. IT-756-13), the Spanish Ministerio de Ciencia e
Innovaci\'{o}n (Grant No. FIS2010-19609-C02-01), the Ministry of
Education and Science of Russian Federation (Grant No. 2.8575.2013),
and the Russian Foundation of Basic Researches (Grant
No.13-02-00016-a).
\end{ack}


\end{document}